\documentclass[aps, prb, twocolumn, showpacs, floatfix, amsmath, amssymb, reprint]{revtex4-1}

\usepackage{graphicx}
\usepackage{dcolumn}			
\usepackage{bm}				
\usepackage{psfrag}
\usepackage{natbib}

\usepackage{amssymb,amsmath}
\usepackage{epstopdf}
\usepackage{epsfig}
\begin{document}

\title{Determination of total x-ray absorption coefficient using non-resonant x-ray emission}
\author{A. J. Achkar$^1$, T. Z. Regier$^2$, E. J. Monkman$^3$, K. M. Shen$^{3,4}$ and D. G. Hawthorn*$^1$ }
\affiliation{$^1$Department of Physics and Astronomy, University of Waterloo, Waterloo, N2L 3G1, Canada \\ $^2$Canadian Light Source, University of Saskatchewan, Saskatoon, Saskatchewan S7N 0X4, Canada \\ $^3$Laboratory of Atomic and Solid State Physics, Department of Physics, Cornell University, Ithaca, NY 14853 \\ $^4$Kavli Institute at Cornell for Nanoscale Science, Cornell University, Ithaca, NY 14853}

\date{\today}

\begin{abstract}
An alternative measure of x-ray absorption spectroscopy (XAS) called inverse partial fluorescence yield (IPFY) has recently been developed that is both bulk sensitive and free of saturation effects. Here we show that the angle dependence of IPFY can provide a measure directly proportional to the total x-ray absorption coefficient, $\mu(E)$. In contrast, fluorescence yield (FY) and electron yield (EY) spectra are offset and/or distorted from $\mu(E)$ by an unknown and difficult to measure amount. Moreover, our measurement can determine $\mu(E)$ in absolute units with no free parameters by scaling to $\mu(E)$ at the non-resonant emission energy. We demonstrate this technique with measurements on NiO and NdGaO$_3$. Determining $\mu(E)$ across edge-steps enables the use of XAS as a non-destructive measure of material composition. In NdGaO$_3$, we also demonstrate the utility of IPFY for insulating samples, where neither EY or FY provide reliable spectra due to sample charging and self-absorption effects, respectively.
\end{abstract}


\maketitle


X-ray absorption spectroscopy (XAS) is widely used in biology, the physical sciences and materials engineering as a powerful probe of spatial and electronic structure.\cite{Lee81,Wende04,Stohr96,deGroot08}  In XAS, the by-products of the absorption process, electron yield (EY) and fluorescence yield (FY), are commonly used as measures of the x-ray absorption\cite{Gudat72,Jaklevic77} since transmission experiments often require impractically thin samples. The principle behind EY and conventional FY measurements (which measure the fluorescence from resonant emission processes and shall henceforth be simply referred to as FY) is that the electron and fluorescence yields bear some proportionality to the absorption coefficient -- the number of electrons or photons emitted from decaying atoms in a given thickness of sample is proportional to the number atoms that are excited. However, the measured FY or EY spectra are not strictly proportional to the total absorption coefficient for several reasons. 

First, the thickness of sample probed depends on the relative penetration depth (attenuation length) of the incident photons and the escape depth of the emitted electrons, in the case of electron yield, or photons, in the case of fluorescence yield. As the attenuation length varies over an absorption edge, it is possible for the attenuation length to approach the electron escape depth, leading to saturation effects in EY and distorting the measured spectra.\cite{Nakajima99} In the case of FY measurements of concentrated species, both the total x-ray absorption coefficient and the absorption due to the edge of interest vary strongly, leading to distortions of the spectra referred to as saturation effects or as ``self-absorption effects.''\cite{Troger92,Eisebitt93} Such FY spectra can sometimes be corrected for self-absorption effects using the angle dependence of the FY.\cite{Troger92,Eisebitt93} However, this correction procedure can be unreliable since resonant x-ray emission processes\cite{KotaniRMP} that are not accounted for in the self-absorption correction can have a significant influence on the energy dependence of the fluorescence yield.\cite{deGroot94b} 

Second, the magnitude of the EY and FY both depend on the relative probability, $\omega_{fl}$, that an excited atom will decay by emitting photons as opposed to electrons.\cite{Hubbell94} This relative probability differs from atom to atom and edge to edge and is generally not known with great precision. 

Third, the emission is distributed over a range of electron and photon energies. A given detector will not detect all electron or photon energies with equal efficiency. In the case of EY, magnetic or electrostatic fields will also influence the efficiency of detection in the system, which may vary between experiments. In addition, the quantum efficiency of EY (the number of electrons emitted per incident photon) will also vary with photon energy.\cite{Stohr96} The consequence of all these factors is that the magnitude of the EY or FY signal will generally have a value that is not proportional to the total absorption coefficient but is rather offset or distorted by some often unknown or difficult to calculate factors.

Fortunately, for many applications of XAS, the key features in absorption spectra measured by EY or FY are retained and can still be interpreted to glean important qualitative information about the electronic or spatial structure. However, in many instances, such as correcting for self-absorption effects, calculating resonant scattering cross-sections or determining x-ray penetration depth, it is important to know the magnitude of the total absorption coefficient in absolute units. Moreover, knowing this could open the door to using XAS as a quantitative tool for compositional analysis of materials. In principle, the magnitude and energy dependence of the total absorption coefficient contains information about the composition of a material in addition to information about the electronic and spatial structure. As the photon energy is increased through an absorption edge, the absorption increases in a step-wise fashion when core electrons are photo-excited with enough energy to enter the continuum of unoccupied states. The magnitude of the edge-step relative to the pre-edge can provide a measure of material composition. The various atomic contributions can be determined using tabulated\cite{Henke93} or calculated\cite{Chantler95} values of the absorption cross-section that are conveniently and freely available online from the Center for X-ray Optics (CXRO) or the National Institute of Standards and Technology (NIST). 
 
With these inputs, the magnitude of the XAS, in particular the edge-step, can be used as a robust quantitative measure of material composition. By fitting the available tabulated or calculated atomic absorption data to the pre- and post-edge of a measured absorption spectrum, one can experimentally derive the stoichiometry of a material in a non-destructive manner. Since they do not measure the total absorption coefficient, FY and EY are not well suited for this type of analysis. Transmission measurements, however, do provide a direct and quantitative measure of the absorption cross-section and such measurements are routinely performed at hard and soft x-ray beamlines.\cite{deGroot10} However, transmission spectra can be subject to ``thickness effects'' and should only be performed with sufficiently thin samples.\cite{Parratt57,Stern81} Preparing samples with appropriate thickness may be challenging or impossible depending on the nature of the sample, particularly for soft x-rays where sample thicknesses less than 1 micron are typically required. 

The recent development of inverse partial fluorescence yield allows us to overcome the aforementioned shortcomings of EY and FY.\cite{Achkar11} Unlike EY and FY measurements, IPFY is both bulk sensitive and free of saturation effects. In this paper, we demonstrate that the theory of IPFY can be extended and exploited to reliably obtain a measure proportional to the total x-ray absorption coefficient, $\mu(E)$, with the proportionality constant being the total absorption coefficient at the non-resonant emission energy, $\mu(E_f)$. This result is confirmed by excellent agreement with tabulated or calculated values of the measured IPFY of NiO and NdGaO$_3$ single crystals. The ability to derive quantitative information from XAS with IPFY creates new opportunities for chemical speciation and compositional analysis of materials. 

In addition, we demonstrate the applicability of IPFY to XAS measurements of strongly insulating samples. In NdGaO$_3$, neither EY or FY measurements provide reliable XAS spectra of the Nd $M_{5,4}$ edges due to strong charging and saturation effects, respectively. In contrast, IPFY provides excellent agreement with previously measured XAS on Nd metal. 

\section*{Results}
\subsection{Inverse partial fluorescence yield}

IPFY operates on a different principle than EY or FY, effectively measuring the attenuation length into a sample rather than the number of atoms that are excited and subsequently relax. With IPFY, an energy sensitive detector is used to monitor non-resonant x-ray emission as the incident photon energy is scanned through an absorption edge. This non-resonant (normal) emission may be from a different element or core electron than that associated with the absorption edge under investigation. As the attenuation length decreases through an absorption edge, the same number of atoms are excited (since all photons are absorbed for samples which are thick relative to the x-ray penetration length), but fewer of these excitations will correspond to non-resonant transitions. Subsequently, the intensity of the non-resonant emission will dip as the absorption coefficient peaks through an absorption edge. 

The intensity of the non-resonant emission will also depend on the absorption cross-section of the atom and core electron corresponding to the non-resonant transition and on the attenuation length of the emitted photons. However, these factors are constant or vary weakly through an absorption edge, in the x-ray absorption near edge structure (XANES). As a result, a straightforward inversion of the partial fluorescence yield (PFY) arising from a non-resonant emission process provides an accurate measure of x-ray absorption cross-section in the XANES.\cite{Achkar11} As discussed in Ref. \citenum{Achkar11}, since it is non-resonant emission processes that contribute to this measure of PFY, saturation (self-absorption) effects are avoided. Moreover, the large variation of the fluorescence decay rates observed across edge steps for resonant fluorescence processes,\cite{deGroot94b} as in conventional FY, do not factor into the measurement of IPFY, simplifying the analysis and interpretation of IPFY relative to FY.

\begin{figure}[htb]
\begin{center}
\resizebox{3.2 in}{!}{\includegraphics{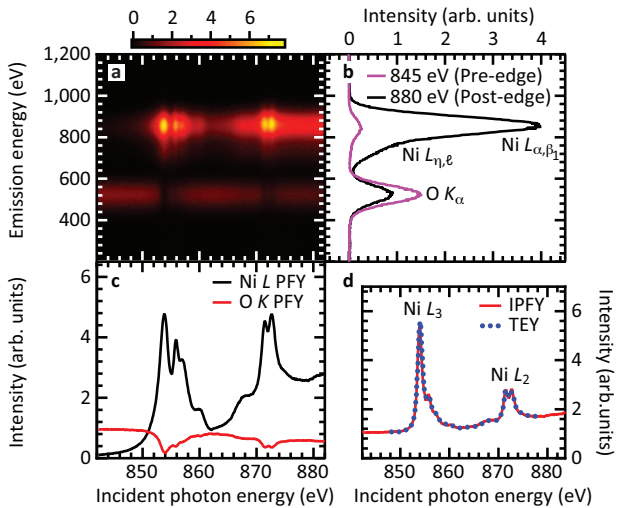}}
\caption{\textbf{Energy sensitive fluorescence yield of NiO} -- \textbf{a)} Normalized x-ray fluorescence of NiO as the incident photon energy is scanned through the Ni $L_3$ and $L_2$ edges. \textbf{b)} The emission spectra in the pre- and post-edge regions at incident photon energies of 845 eV and 880 eV taken in 1-eV windows. Emissions corresponding to the resonant Ni $3d$ to $2p$ ($L_{\alpha,\beta_1}$) and $3s$ to $2p$ ($L_{\eta,\ell}$) and non-resonant (normal) O $2p$ to $1s$ ($K_\alpha$) processes are observed. \textbf{c)} The Ni $L$ and O $K$ partial fluorescence yield extracted from panel a in 150-eV wide energy windows centered on the respective emissions. The resonant Ni $L$ PFY shows strong distortions resulting from saturation effects. The normal O $K$ PFY dips as the absorption increases through the Ni $L_{3,2}$ absorption edges. \textbf{d)} The IPFY is the inverse of the O $K$ PFY shown in panel c. The NiO IPFY is in good agreement with total electron yield data from Ref.~\citenum{Abbate91} which has been scaled and offset to match the IPFY. \label{fig:fig1} }
\end{center}
\end{figure}

The extraction of IPFY from the energy-resolved x-ray emission of NiO is demonstrated in Fig.~\ref{fig:fig1}. The x-ray emission of NiO is measured as the incident photon energy, $E_i$, is scanned through the Ni $L$ edge (Fig.~\ref{fig:fig1}a). The Ni $L$ absorption edge corresponds to exciting a Ni $2p$ electron into unoccupied $3d$ states near the edge (and a continuum of states further above the edge), leaving behind a $2p$ core hole. The emission spectra (Fig.~\ref{fig:fig1}b) exhibit a peak at emission energy $E_f \sim$ 840 eV that corresponds to resonant emission from Ni. This emission is due to the electrons making transitions to fill in the Ni $2p$ core-hole left behind by the Ni $L$ edge absorption process. The PFY from the Ni $2p$ emission (Fig.~\ref{fig:fig1}c, black curve) suffers significantly from self-absorption effects and bears little resemblance to the absorption coefficient. 

In addition to the Ni $L$ absorption, the x-ray absorption and emission also have contributions from non-resonant transitions of other core electrons of Ni ($3s$, $3p$) and from oxygen (the total linear absorption coefficient is the sum of these contributions, $\mu(E_i) = \mu_{Ni}(E_i) + \mu_O(E_i)$, where $\mu_{Ni}(E_i) = \mu_{Ni,2p}(E_i) + \mu_{Ni,3s}(E_i) + \mu_{Ni,3p}(E_i) + $ \textellipsis).\cite{Yeh85} As shown in Fig.~\ref{fig:fig1}a and 1b there is a band of emission centred at 524 eV corresponding to the non-resonant emission of O $2p$ valence electrons decaying to fill in the O $1s$ core hole (O $K$ emission). The PFY from the O $K$ emission (Fig.~\ref{fig:fig1}c, red curve) exhibits dips at the Ni $L_{3,2}$ absorption edges. The inverse of this spectrum, the IPFY = 1/PFY$_{\text{O} K}$, is shown in Fig.~\ref{fig:fig1}d along with total electron yield (TEY) measurements of NiO from Ref.~\citenum{Abbate91} that have been scaled and offset to match the IPFY. Similar to previous work\cite{Achkar11} on La$_{1.475}$Nd$_{0.4}$Sr$_{0.125}$CuO$_4$, the agreement between IPFY and TEY is very good, highlighting the ability of IPFY to measure the energy dependence of the absorption coefficient of Ni without the strong self-absorption effects experienced with PFY.

\subsection{Geometry factor of IPFY in the XANES region}
It has been shown that the IPFY of thick, homogeneous materials is a function of the total x-ray absorption coefficient $\mu(E_i$):\cite{Achkar11}
\begin{equation} \label{eq:eqn1}
IPFY=\frac{I_0(E_i)}{I(E_i,E_f)}=A\left(\mu(E_i)+B\right)
\end{equation}
where $A=4\pi/\eta(E_f) \Omega \omega_Y(E_f)\mu_Y(E_i)$ and  $B=\mu(E_f)\frac{\sin\alpha}{\sin\beta}$. Here $\alpha$ and $\beta$ are the angles of incidence and emission, respectively, as measured from the sample surface, $\eta(E_f)$ is the quantum efficiency of the detector at the emission energy, $\Omega$ is the detector solid angle,  $\mu_Y(E_i)$ is the contribution to the total absorption coefficient from the excitation of core electron $Y$ (ex. O $1s$) and $\omega_Y(E_f)$ is the probability of fluorescence at energy $E_f$ resulting from electrons decaying to fill in the core hole left by $Y$. 
\begin{figure}[ht]
\begin{center}
\resizebox{3.2in}{!}{\includegraphics{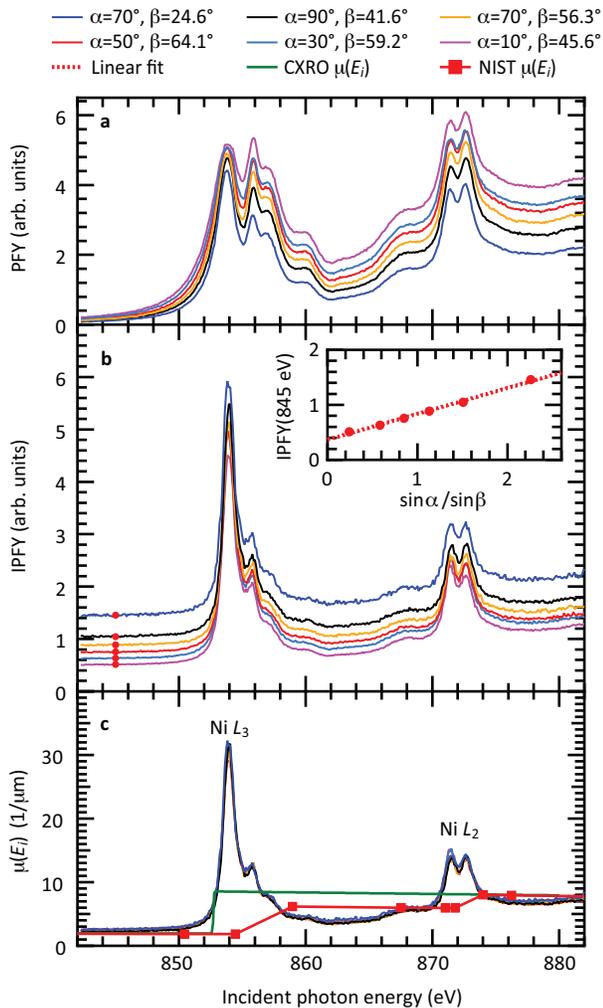}}
\caption{\textbf{Angle dependence of PFY and IPFY} -- \textbf{a)} The Ni $L$ PFY for various experimental geometries. The spectra are distorted by strong self-absorption effects that depend on the angle of incidence ($\alpha$) and angle of emission ($\beta$). \textbf{b)} The IPFY extracted from the O $K$ PFY for the same experimental geometries as panel a. The spectra are offset by a geometry dependent constant, but are otherwise not distorted. The inset plots the IPFY at $E_i = 845$ eV (red circles) as a function of $\sin\alpha/\sin\beta$, which varies linearly as predicted by Eq.~\eqref{eq:eqn1}. \textbf{c)} The linear absorption coefficient, $\mu(E_i)$, obtained from IPFY spectra. As described in the text, the offsets in the IPFY spectra are subtracted, collapsing the IPFY spectra onto a single curve proportional to $\mu(E_i)$. The spectra shown here have been scaled using a single tabulated\cite{Henke93} value for $\mu(E_f)$ and plotted against the tabulated\cite{Henke93} (green) and calculated\cite{Chantler95} (red squares) absorption coefficients. \label{fig:fig2}}
\end{center}
\end{figure}

In Eq.~\eqref{eq:eqn1}, the constant $B$ is independent of $E_i$ and $A$ depends only weakly on $E_i$ over a narrow energy range (XANES) so it can be treated approximately as constant.\cite{Achkar11} This approximation fails over a large energy range, requiring one to account for the energy dependence of $\mu_Y(E_i)$ and the quantum efficiency of the $I_0$ measurement, which we demonstrate later. However, in a narrow energy range, it follows that IPFY is proportional to $\mu(E_i)$ plus an offset proportional to $B$. The crucial feature of Eq.~\eqref{eq:eqn1} is that the size of the offset $B$ is determined by the geometrical factor $\sin\alpha / \sin \beta$. This allows one to determine $\mu(E_i)/\mu(E_f)$ from experiments with different measurement geometries.

In Fig.~\ref{fig:fig2}, we demonstrate that the IPFY of NiO obeys the expected dependence on the sample geometry as detailed in Eq.~\eqref{eq:eqn1}. First, the Ni $L_{3,2}$ PFY spectra measured for various geometries (Fig.~\ref{fig:fig2}a) depict the strong angle-dependence of self-absorption effects in FY measurements. Notably, attempts to correct the PFY for self-absorption effects using the angle dependence\cite{Troger92,Eisebitt93} (not shown) do not yield the correct spectra. In contrast, the IPFY spectra measured with the same geometries (Fig.~\ref{fig:fig2}b) are undistorted and offset from one another, in agreement with Eq.~\eqref{eq:eqn1}. The inset in Fig.~\ref{fig:fig2}b is a plot of the value of the IPFY spectra at a single value of the incident photon energy [$E_i=845$ eV (red circles)] as a function of  $\sin\alpha / \sin \beta$ for the given experimental geometries. As expected, this offset fits well to a straight line with an intercept equal to $A\mu(845$ eV) and a slope equal to $A\mu(E_f)$. Subtracting $A\mu(E_f) \sin\alpha / \sin \beta$ for each of the spectra, we find that they collapse onto a single curve (the slight variations in peak intensities are primarily due to magnetic linear dichroism in NiO due to anti-ferromagnetic ordering of the Ni spins in the $(111)$ plane\cite{Stohr98}). The key point of this analysis is that the resulting spectra, derived entirely from experiment, are directly proportional to the total absorption coefficient without any offsets. 

The proportionality to $\mu(E_i)$ is verified by comparing our measurement to tabulated\cite{Henke93} and calculated\cite{Chantler95} values of $\mu(E_i)$. The calculated and tabulated data capture the transitions from the core electron to the continuum, accurately reproducing the edge-step, but do not include the multiplet physics associated with the $2p$ to $3d$ transition. We use the calculated value of the absorption coefficient at the O $K$ emission energy ($\mu(E_f =524 \text{eV}) = 3.14\times10^6$ m$^{-1}$ for NiO from Ref.~\citenum{Henke93}) to normalize the subtracted offset and determine the proportionality constant $A$. Note that the O $K$ emission is due primarily to $2p$ valence electrons decaying to fill the $1s$ core hole and is peaked at a photon energy below the absorption threshold. The data shown in Fig.~\ref{fig:fig2}c has been scaled using $\mu(E_f)$ (a non-arbitrary scaling factor) and is shown along with the tabulated\cite{Henke93} (green curve) and calculated\cite{Chantler95} (red squares) x-ray absorption coefficient. 

Using this single scaling parameter, we find that the measured spectra are in excellent agreement with the tabulated coefficients in both the pre- and post-edge regions, capturing both the energy dependence and the edge-step. This demonstrates that IPFY provides a measure directly proportional to the total absorption coefficient with the proportionality constant being $\mu(E_f)$. In contrast, quantitative analysis of EY or FY spectra requires scaling and offsetting data to calculated values of the absorption coefficient above and below the edge, essentially fixing the edge-step.\cite{Stohr96} This latter procedure requires prior knowledge of the material composition and is subject to uncertainties in the tabulated or calculated values which are estimated at 5-20\% between 500 and 1000 eV and even higher near absorption edges.\cite{Chantler00} Moreover, XAS measurements often still have significant structure above an absorption edge in the form of extended x-ray absorption fine structure (EXAFS) that is not accounted for in the tabulated or calculated values, resulting in additional errors in normalizing data above an absorption edge. In contrast, with IPFY, we obtain the energy dependence of $\mu(E_i)$ directly from measurement and can scale the data at a single point well below the absorption edge. The result of this normalization can be independently checked against the absorption above and below the absorption edge in question and multiple angles can be measured to ensure self-consistency, resulting in a reliable and accurate normalization of the data.
 
\subsection{IPFY beyond the XANES}

\begin{figure}[h!t]
\begin{center}
\resizebox{3.2in}{!}{\includegraphics{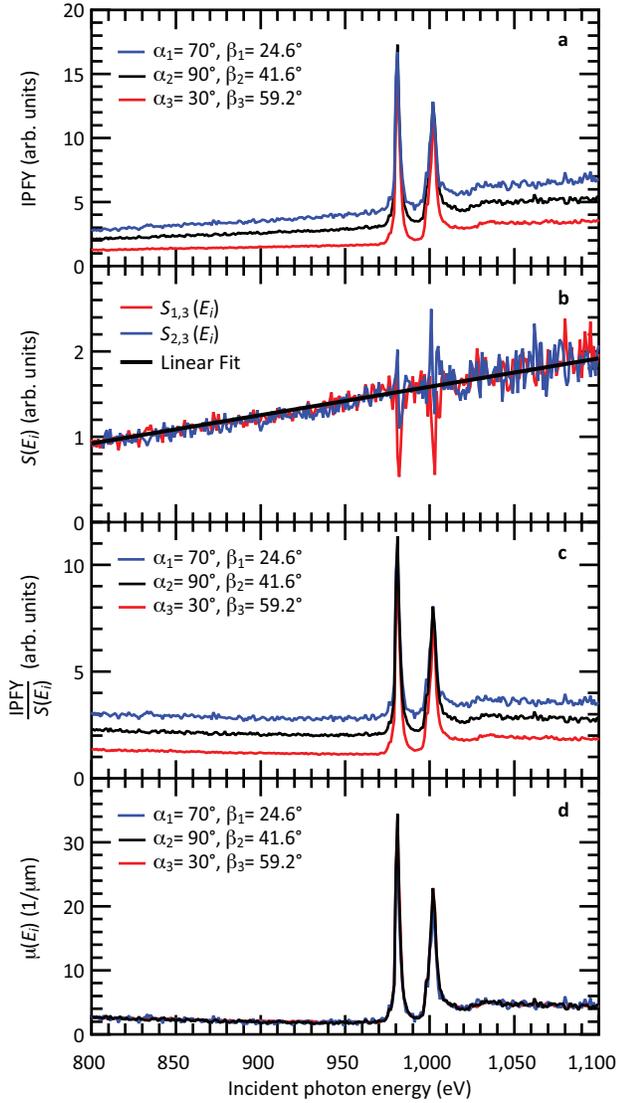}}
\caption{\textbf{Wide energy range IPFY of NdGaO$_{\boldsymbol 3}$} -- \textbf{a)} IPFY of NdGaO$_3$ for several measurement geometries. The IPFY is measured using the O $K$ emission in a 150 eV window centred about 524 eV. The measurements at different geometries exhibit different sloping backgrounds due to the energy dependence of $\mu_{\text{O} K}(E_i)$ and the quantum efficiency of the $I_0$ measurement, $\nu(E_i)$. \textbf{b)} $S(E_i)$ calculated using Eq.~\eqref{eq:eqnB2} with the different measurement geometries depicted in the legend of panel a. The black line is a linear fit to $S(E_i)$. \textbf{c)} The IPFY$(E_i)/S(E_i)$ spectra are rigidly offset by $B$. \textbf{d)} The total absorption coefficient, $\mu(E_i)$, determined using Eq.~\eqref{eq:eqnB4} (the data are scaled to $\mu$(524 eV) from Ref.~\citenum{Chantler95}). The spectra measured with different geometries collapse onto a single curve over the entire energy range. \label{fig:fig3}}
\end{center}
\end{figure}

In the NiO measurements shown above, the described offsetting procedure works well over the narrow energy range covered, giving a quantity approximately proportional to $\mu(E_i)$. However, over a larger energy window, the energy dependence of $\mu_Y(E_i)$ can be significant. An example of this effect is shown in measurements of NdGaO$_3$ over a wide energy range. In Fig.~\ref{fig:fig3}a, the IPFY measured using the O $K$ emission of NdGaO$_3$ is shown for three measurement geometries over an extended energy range covering the Nd $M$ edge. 

The spectra are not rigidly offset, instead appearing to be subject to a sloping background in addition to an offset. This background is due to the energy dependence of $\mu_{\text{O} K}(E_i)$ and also to the energy dependence of our measurement of the incident photon flux, $I_0$. 

In our measurement, and many XAS measurements, $I_0$ is measured using a Au grid with 85\% transmission that is placed between the sample and the last optical component. The total electron yield from the grid, $I_{\text{Grid}}$, is used to measure the incident photon flux. This measurement, however, depends not only on $I_0$, but also on the quantum efficiency of the mesh, $\nu (E_i)$ (the number of electrons generated per incident photon), which in general will be energy dependent. As such, $I_{\text{Grid}}(E_i)  = I_0(E_i) \nu (E_i)$ and Eq.~\eqref{eq:eqn1} should be modified to:
\begin{equation} \label{eq:eqnB1}
\begin{split}
IPFY=\frac{I_{\text{Grid}}(E_i)}{I(E_i,E_f)} &=\frac{I_0(E_i)\nu(E_i)}{I(E_i,E_f)} \\
& \approx  \frac{D \nu(E_i)}{\mu_Y(E_i)} \left(\mu(E_i)+B\right)
\end{split}
\end{equation}
where $D = A \mu_Y(E_i)$. Fortunately, the energy dependence of both $\nu(E_i)$ and $\mu_Y(E_i)$ can be unambiguously eliminated from the data by subtracting IPFY spectra measured with different measurement geometries and normalizing to the geometry ($\nu(E_i)$ generally also enters into EY and FY measurements, but is typically not corrected for). From Eq.~\eqref{eq:eqnB1} it follows that
\begin{equation} \label{eq:eqnB2}
\begin{split}
S_{j,k}(E_i)&=\frac{D\nu(E_i)}{\mu_Y(E_i)}\mu(E_f)\\
&=\frac{IPFY(\alpha_j,\beta_j)-IPFY(\alpha_k,\beta_k)}{\frac{\sin\alpha_j}{\sin\beta_j}-\frac{\sin\alpha_k}{\sin\beta_k}}
\end{split}
\end{equation}
where $j$ and $k$ correspond to different measurement geometries and $S(E_i)$ is independent of the choice of $j$ and $k$. We can now write 
\begin{equation} \label{eq:eqnB3}
\frac{IPFY}{S(E_i)} = \frac{1}{\mu(E_f)} \left( \mu(E_i)+\mu(E_f)\frac{\sin\alpha}{\sin\beta}\right),
\end{equation} 
which is simply rearranged to yield the total x-ray absorption coefficient:
\begin{equation}  \label{eq:eqnB4}
\mu(E_i)=\mu(E_f)\left( \frac{IPFY}{S(E_i)}-\frac{\sin\alpha}{\sin\beta} \right).
\end{equation}

\begin{figure}[htb!]
\begin{center}
\resizebox{3.2 in}{!}{\includegraphics{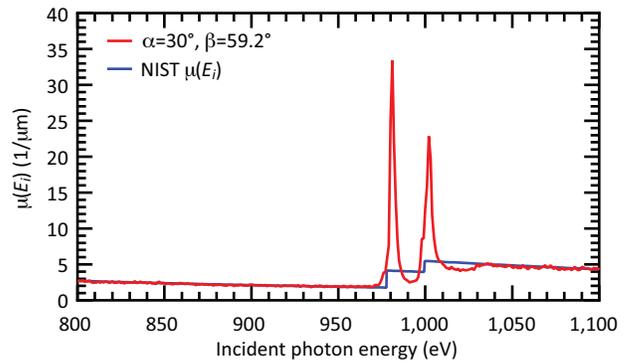}}
\caption{\textbf{Normalized IPFY compared to atomic calculations} -- The absorption coefficient of NdGaO$_3$ extracted from the O $K$ IPFY and corrected for the energy dependence of the O $K$ absorption and the quantum efficiency of the $I_0$ measurement. The incident photon energy was scanned across the Nd $M_5$ and $M_4$ edges. The data agrees well with calculated XAS\cite{Chantler95} over a wide energy range. \label{fig:fig4}}
\end{center}
\end{figure}

In Fig.~\ref{fig:fig3}, this subtraction is shown, giving $S(E_i)$ that is a smooth function of energy. As shown in Fig.~\ref{fig:fig3}c, dividing the spectra in Fig.~\ref{fig:fig3}a by $S(E_i)$, provides spectra that are rigidly offset over a wide range in energy. Subtracting $\sin\alpha / \sin\beta$ from the spectra provides $\mu(E_i)/\mu(E_f)$, collapsing the data onto a single curve, which is then scaled using a calculated value\cite{Chantler95} of $\mu(E_f=524~\text{eV})$ as shown in Fig.~\ref{fig:fig3}d. When normalized in this way, the spectra are in excellent quantitative agreement with the calculated absorption coefficient over a wide energy range above and below the Nd $M_{5,4}$ absorption edge, as shown in Fig.~\ref{fig:fig4}. 

\subsection{IPFY in strong insulators}
\begin{figure}[htb!]
\begin{center}
\resizebox{3.2 in}{!}{\includegraphics{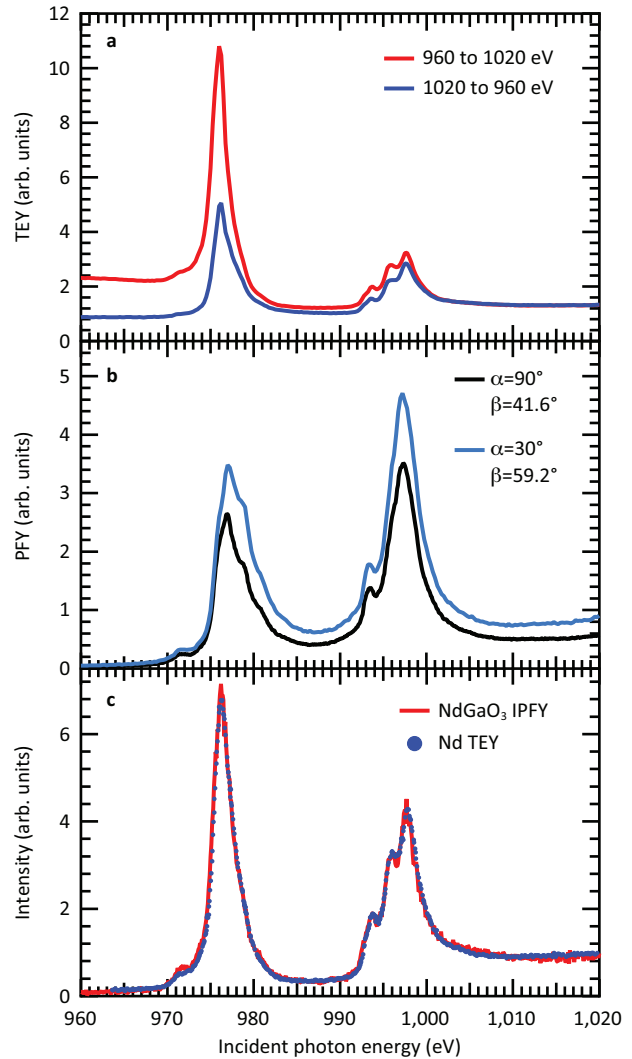}}
\caption{ \textbf{XAS of NdGaO$_{\boldsymbol 3}$} -- \textbf{a)} The TEY of NdGaO$_3$ exhibits an anomalous negative edge-jump across the Nd $M_{5,4}$ edges (red curve). A spectrum collected with the incident photon energy scanned in the negative direction (blue curve) soon after has positive edge-jumps. This difference is attributed to a charge up of the sample surface, affecting the TEY measurement. Neither spectrum matches well with TEY on pure metallic Nd from Ref.~\citenum{Thole85}. \textbf{b)} The partial fluorescence yield from the Nd emission of NdGaO$_3$ is strongly distorted by saturation effects. \textbf{c)} The IPFY extracted from the O $K$ PFY of NdGaO$_3$ agrees remarkably well with the TEY of pure Nd from Ref.~\citenum{Thole85} which is scaled and offset to match the IPFY. \label{fig:Fig5}}
\end{center}
\end{figure}

Finally, we would like to emphasize the role of IPFY to study insulating samples that can be difficult or impossible to measure correctly using FY or EY.  An example of such a system is NdGaO$_3$.  This material is an insulator commonly used as a substrate for oxide film growth.  EY measurements of the Nd $M$ edge in NdGaO$_3$, shown in Fig.~\ref{fig:Fig5}a, exemplify issues one can encounter when measuring the EY of samples. Here the EY has an unphysical negative edge jump at the absorption edge.  The unusual behaviour is attributed to a build-up in positive charge near the surface of the sample that effectively reduces the number of emitted electrons. We were able to reduce the effect by recording the spectra by scanning the incident photon energy in the negative direction (1020 eV to 980 eV) or measuring different spots on the sample, but ultimately these spectra are not reliable.  

PFY and TFY in this material are also unreliable.  The Nd edge PFY measurements, shown in Fig.~\ref{fig:Fig5}b, are heavily distorted by self-absorption effects, similar to NiO.  In contrast, the IPFY (Fig.~\ref{fig:Fig5}c) provides the correct XAS spectrum for Nd$^{3+}$.  This is evidenced by excellent agreement with XAS in pure Nd, which like NdGaO$_3$ has Nd$^{3+}$ character and is described well by atomic multiplet calculations.\cite{Thole85}  In this case, both EY and FY provide erroneous results and transmission measurements are not possible due to the thickness of the sample.  As such, IPFY provides the only means to measure the correct XAS spectrum. We anticipate IPFY to be widely applicable to similar cases.

\section*{Discussion}
Experimental studies that require accurate knowledge of optical constants or atomic scattering form factors, such as in modelling of resonant reflectivity or x-ray scattering, stand to benefit substantially from angle dependent IPFY since it provides a measure of the total absorption coefficient. In such studies, it is common to scale and offset XAS spectra above and below an absorption edge to tabulated atomic calculations or absorption data.\cite{Stohr06} This procedure requires knowledge of the composition of a material and requires measurements that extend sufficiently above absorption edges to avoid EXAFS resonances. It is not always possible to meet these requirements, and in such cases the determination of optical constants or atomic scattering form factors will necessarily be subject to systematic errors. In contrast, with angle dependent IPFY, $\mu(E_i)$ and $\mu(E_f)$ can be determined with a simple fitting approach that does not depend on prior knowledge of material composition. Consequently, scaling the measured absorption to absolute units using $\mu(E_f)$ enables the determination of atomic form factors with the appropriate edge-step even if sample composition is not previously known or if the XAS spectra do not extend sufficiently above the EXAFS. 

As an accurate measure of $\mu(E_i)$, IPFY spectroscopy could become a powerful tool in non-destructive quantitative analysis of material composition, which can be done separately or in conjunction with XANES or EXAFS measurements of electronic and spatial structure. Without prior knowledge of material composition, it is possible to fit $\mu(E_i)/\mu(E_f)$ to a sum of the tabulated atomic absorption coefficients in order to determine the relative weights of each atomic species in a sample. Furthermore, $\mu(E_f)$ can be determined by the fitting routine as it too is the weighted sum of the atomic contributions. Thus, in a fully self-consistent way, it is possible to utilize IPFY spectra to estimate the composition of an unknown sample. Even if a quantitative estimate is not needed, the magnitude of the pre-edge relative to the post-edge bears a distinct signature of the quantity of an element relative to the other elements in the material. A simple comparison of the magnitude of the edge-step compared to calculations or to IPFY on similar materials can then be used as a clear measure of sample composition. We believe this kind of non-destructive estimate of sample composition will be very useful to XAS practitioners as a simple means to verify the stoichiometry of a given sample.

Finally, we would like to comment on the applicability of IPFY to the hard x-ray regime. Thus far, IPFY has only been demonstrated using soft x-rays. However, we feel IPFY would likely also be useful for XAS at hard x-ray energies. In order to measure IPFY in this case, one would require the appropriate selection of emission lines. While low energy emission lines would exist, their excitation cross-section would be quite small and the presence of air and/or windows between the sample and the detector may make it impossible to detect these. However, in compounds with multiple elements, one could in principle utilize non-resonant $K$ or $L$ emission lines (at intermediate to hard x-ray energies) to study the $K$ edge absorption of another element. Hence, we believe that IPFY studies at hard x-ray energies are feasible and could be performed using a similar detection scheme as we have used at soft x-ray energies. 

In conclusion, we have demonstrated a measure of the total x-ray absorption coefficient using angle dependent IPFY. Unlike in EY or conventional FY measurements, the offset in IPFY can be subtracted unambiguously from experiments with varied geometry to provide data directly proportional to $\mu(E_i)$ and undistorted by saturation or self-absorption effects. By scaling to a single value of $\mu(E_f)$, $\mu(E)$ is obtained in absolute units. We anticipate this technique to have wide applicability in many areas of science and engineering, potentially opening XAS up to non-destructive, quantitative analysis of material composition. 

\section*{Methods}
The XAS measurements were performed at the Canadian Light Source's 11-ID SGM beamline. All measurements were made at room temperature. The drain current of the sample provided the electron yield. An energy-dispersive silicon drift detector (SDD) with an energy resolution of $\sim$120 eV was used to collect the emission spectra as a function of incident photon energy. The SDD was fixed in position ($25.8^{\circ}$ below the plane and $42.5^{\circ}$ from the beam axis) and the sample was rotated about the vertical axis to vary $\alpha$ and $\beta$, the angles of incidence and emission, respectively. Dark counts on the detector were negligible. However, a small background in the 200-2000 eV region of the NiO emission spectra was observed, likely due to a slight mis-calibration of the detector electronics. This background potentially introduced an error of up to 20\% at the Ni $L_3$ peak and 3\% in the post-edge.

The single crystal of cubic NiO was polished to a surface roughness less than 0.03 $\mu$m. Its dimensions were 5$\times$5 by 0.5 mm thick and it was oriented such that the $\langle 100 \rangle$ direction was perpendicular to the sample surface. The NdGaO$_3$ single crystal was a 10$\times$10 mm by 0.5 mm thick, polished substrate oriented with the $\langle 100 \rangle$ direction perpendicular to the sample surface. 
\\ \\

\begin{acknowledgements}
This research is supported by the Natural Sciences and Engineering Research Council of Canada and by the National Science Foundation through DMR-0847385 and the MRSEC program under DMR-0520404 (Cornell Center for Materials Research). The research described in this paper was performed at the Canadian Light Source, which is supported by NSERC, NRC, CIHR, and the University of Saskatchewan. E.J.M. acknowledges NSERC for PGS support.
\end{acknowledgements}


\begin{thebibliography}{10}
\expandafter\ifx\csname url\endcsname\relax
  \def\url#1{\texttt{#1}}\fi
\expandafter\ifx\csname urlprefix\endcsname\relax\def\urlprefix{URL }\fi
\providecommand{\bibinfo}[2]{#2}
\providecommand{\eprint}[2][]{\url{#2}}

\bibitem{Lee81}
\bibinfo{author}{Lee, P.}, \bibinfo{author}{Citrin, P.},
  \bibinfo{author}{Eisenberger, P.} \& \bibinfo{author}{Kincaid, B.}
\newblock Extended X-ray absorption fine structure - its strengths and
  limitations as a structural tool.
\newblock \emph{\bibinfo{journal}{Rev. Mod. Phys}}
  \textbf{\bibinfo{volume}{53}}, \bibinfo{pages}{769} (\bibinfo{year}{1981}).

\bibitem{Wende04}
\bibinfo{author}{Wende, H.}
\newblock Recent advances in x-ray absorption spectroscopy.
\newblock \emph{\bibinfo{journal}{Rep. Prog. Phys.}}
  \textbf{\bibinfo{volume}{67}}, \bibinfo{pages}{2105--2181}
  (\bibinfo{year}{2004}).

\bibitem{Stohr96}
\bibinfo{author}{St\"ohr, J.}
\newblock \emph{\bibinfo{title}{NEXAFS Spectroscopy}}
  (\bibinfo{publisher}{Springer}, \bibinfo{address}{New York},
  \bibinfo{year}{1996}).

\bibitem{deGroot08}
\bibinfo{author}{deGroot, F.} \& \bibinfo{author}{Kotani, A.}
\newblock \emph{\bibinfo{title}{Core Level Spectroscopy of Solids}}
  (\bibinfo{publisher}{CRC Press}, \bibinfo{address}{Boca Raton, FL},
  \bibinfo{year}{2008}).

\bibitem{Gudat72}
\bibinfo{author}{Gudat, W.} \& \bibinfo{author}{Kunz, C.}
\newblock Close similarity between photoelectric yield and photoabsorption
  spectra in the soft-x-ray range.
\newblock \emph{\bibinfo{journal}{Phys. Rev. Lett.}}
  \textbf{\bibinfo{volume}{29}}, \bibinfo{pages}{169--172}
  (\bibinfo{year}{1972}).

\bibitem{Jaklevic77}
\bibinfo{author}{Jaklevic, J.} \emph{et~al.}
\newblock Fluorescence detection of EXAFS: Sensitivity enhancement for dilute
  species and thin films.
\newblock \emph{\bibinfo{journal}{Solid State Commun.}}
  \textbf{\bibinfo{volume}{23}}, \bibinfo{pages}{679--682}
  (\bibinfo{year}{1977}).

\bibitem{Nakajima99}
\bibinfo{author}{Nakajima, R.}, \bibinfo{author}{St\"ohr, J.} \&
  \bibinfo{author}{Idzerda, Y.~U.}
\newblock Electron-yield saturation effects in $L$-edge x-ray magnetic circular
  dichroism spectra of {Fe}, {Co}, and {Ni}.
\newblock \emph{\bibinfo{journal}{Phys. Rev. B}} \textbf{\bibinfo{volume}{59}},
  \bibinfo{pages}{6421} (\bibinfo{year}{1999}).

\bibitem{Troger92}
\bibinfo{author}{Tr\"oger, L.} \emph{et~al.}
\newblock Full correction of the self-absorption in soft-fluorescence extended
  x-ray-absorption fine structure.
\newblock \emph{\bibinfo{journal}{Phys. Rev. B}} \textbf{\bibinfo{volume}{46}},
  \bibinfo{pages}{3283--3289} (\bibinfo{year}{1992}).

\bibitem{Eisebitt93}
\bibinfo{author}{Eisebitt, S.}, \bibinfo{author}{B\"oske, T.},
  \bibinfo{author}{Rubensson, J.-E.} \& \bibinfo{author}{Eberhardt, W.}
\newblock Determination of absorption coefficients for concentrated samples by
  fluorescence detection.
\newblock \emph{\bibinfo{journal}{Phys. Rev. B}} \textbf{\bibinfo{volume}{47}},
  \bibinfo{pages}{14103--14109} (\bibinfo{year}{1993}).

\bibitem{KotaniRMP}
\bibinfo{author}{Kotani, A.} \& \bibinfo{author}{Shin, S.}
\newblock Resonant inelastic x-ray scattering spectra for electrons in solids.
\newblock \emph{\bibinfo{journal}{Rev. Mod. Phys.}}
  \textbf{\bibinfo{volume}{73}}, \bibinfo{pages}{203--246}
  (\bibinfo{year}{2001}).

\bibitem{deGroot94b}
\bibinfo{author}{de~Groot, F.}, \bibinfo{author}{Arrio, M.},
  \bibinfo{author}{Sainctavit, P.}, \bibinfo{author}{Cartier, C.} \&
  \bibinfo{author}{Chen, C.}
\newblock Fluorescence yield detection: Why it does not measure the X-ray
  absorption cross section.
\newblock \emph{\bibinfo{journal}{Solid State Communications}}
  \textbf{\bibinfo{volume}{92}}, \bibinfo{pages}{991--995}
  (\bibinfo{year}{1994}).

\bibitem{Hubbell94}
\bibinfo{author}{Hubbell, J.} \emph{et~al.}
\newblock A review, bibliography, and tabulation of $K$, $L$ and higher atomic
  shell x-ray fluorescence yields.
\newblock \emph{\bibinfo{journal}{J. Phys. Chem. Ref. Data}}
  \textbf{\bibinfo{volume}{23}}, \bibinfo{pages}{339} (\bibinfo{year}{1994}).

\bibitem{Henke93}
\bibinfo{author}{Henke, B.}, \bibinfo{author}{Gullikson, E.} \&
  \bibinfo{author}{Davis, J.}
\newblock X-ray interactions: photoabsorption, scattering, transmission, and
  reflection at {$E$=50-30000 eV, $Z$=1-92}.
\newblock \emph{\bibinfo{journal}{Atomic Data and Nuclear Data Tables}}
  \textbf{\bibinfo{volume}{54}}, \bibinfo{pages}{181} (\bibinfo{year}{1993}).

\bibitem{Chantler95}
\bibinfo{author}{Chantler, C.}
\newblock Theoretical form factor, attenuation, and scattering tabulation for
  $Z$=1-92 from $E$=1-10 eV to $E$=0.4-1.0 MeV.
\newblock \emph{\bibinfo{journal}{J. Phys. Chem. Ref. Data}}
  \textbf{\bibinfo{volume}{24}}, \bibinfo{pages}{71} (\bibinfo{year}{1995}).

\bibitem{deGroot10}
\bibinfo{author}{de~Groot, F. M.~F.}, \bibinfo{author}{de~Smit, E.},
  \bibinfo{author}{van Schooneveld, M.~M.}, \bibinfo{author}{Aramburo, L.~R.}
  \& \bibinfo{author}{Weckhuysen, B.~M.}
\newblock In-situ Scanning Transmission X-Ray Microscopy of Catalytic Solids
  and Related Nanomaterials.
\newblock \emph{\bibinfo{journal}{ChemPhysChem}} \textbf{\bibinfo{volume}{11}},
  \bibinfo{pages}{951--962} (\bibinfo{year}{2010}).

\bibitem{Parratt57}
\bibinfo{author}{Parratt, L.~G.}, \bibinfo{author}{Hempstead, C.~F.} \&
  \bibinfo{author}{Jossem, E.~L.}
\newblock Thickness effect in absorption spectra near absorption edges.
\newblock \emph{\bibinfo{journal}{Phys. Rev}} \textbf{\bibinfo{volume}{105}},
  \bibinfo{pages}{1228--1232} (\bibinfo{year}{1957}).

\bibitem{Stern81}
\bibinfo{author}{Stern, E.~A.} \& \bibinfo{author}{Kim, K.}
\newblock Thickness effect on the extended-x-ray-absorption-fine-structure
  amplitude.
\newblock \emph{\bibinfo{journal}{Physical Review B}}
  \textbf{\bibinfo{volume}{23}}, \bibinfo{pages}{3781} (\bibinfo{year}{1981}).

\bibitem{Achkar11}
\bibinfo{author}{Achkar, A.~J.} \emph{et~al.}
\newblock Bulk sensitive x-ray absorption spectroscopy free of self-absorption
  effects.
\newblock \emph{\bibinfo{journal}{Phys. Rev. B}} \textbf{\bibinfo{volume}{83}},
  \bibinfo{pages}{081106} (\bibinfo{year}{2011}).

\bibitem{Yeh85}
\bibinfo{author}{Yeh, J.} \& \bibinfo{author}{Lindau, I.}
\newblock Atomic subshell photoionization cross sections and asymmetry
  parameters: $1 \leq Z \leq 103$.
\newblock \emph{\bibinfo{journal}{At. Data Nucl. Data Tables}}
  \textbf{\bibinfo{volume}{32}}, \bibinfo{pages}{1} (\bibinfo{year}{1985}).

\bibitem{Abbate91}
\bibinfo{author}{Abbate, M.} \emph{et~al.}
\newblock Soft-x-ray-absorption studies of the location of extra charges
  induced by substitution in controlled-valence materials.
\newblock \emph{\bibinfo{journal}{Phys. Rev. B}} \textbf{\bibinfo{volume}{44}},
  \bibinfo{pages}{5419--5422} (\bibinfo{year}{1991}).

\bibitem{Stohr98}
\bibinfo{author}{St\"ohr, J.}, \bibinfo{author}{Padmore, H.},
  \bibinfo{author}{Anders, S.}, \bibinfo{author}{Stammler, T.} \&
  \bibinfo{author}{Scheinfein, M.}
\newblock Principles of x-ray magnetic dichroism spectromicroscopy.
\newblock \emph{\bibinfo{journal}{Surf. Rev. Lett.}}
  \textbf{\bibinfo{volume}{5}}, \bibinfo{pages}{1297} (\bibinfo{year}{1998}).

\bibitem{Chantler00}
\bibinfo{author}{Chantler, C.}
\newblock Detailed tabulation of atomic form factors, photoelectric absorption
  and scattering cross section, and mass attenuation coefficients in the
  vicinity of absorption edges in the soft x-ray($Z$=30-36, $Z$=60-89, $E$=0.1
  keV-10 keV), addressing convergence issues of earlier work.
\newblock \emph{\bibinfo{journal}{J. Phys. Chem. Ref. Data}}
  \textbf{\bibinfo{volume}{29}}, \bibinfo{pages}{597} (\bibinfo{year}{2000}).

\bibitem{Thole85}
\bibinfo{author}{Thole, B.~T.} \emph{et~al.}
\newblock 3$d$ x-ray-absorption lines and the 3$d^9$4$f^{n+1}$ multiplets of
  the lanthanides.
\newblock \emph{\bibinfo{journal}{Phys. Rev. B}} \textbf{\bibinfo{volume}{32}},
  \bibinfo{pages}{5107--5118} (\bibinfo{year}{1985}).

\bibitem{Stohr06}
\bibinfo{author}{St\"ohr, J.} \& \bibinfo{author}{Seigmann, H.}
\newblock \emph{\bibinfo{title}{Magnetism: From Fundamentals to Nanoscale
  Dynamics.}} (\bibinfo{publisher}{Springer-Verlag},
  \bibinfo{address}{Berlin-Heidelberg}, \bibinfo{year}{2006}).

\end{thebibliography}


\end{document}